\documentclass[12pt, prd, showpacs]{revtex4}
\usepackage{amssymb}
\usepackage{amsmath}

\input{tcilatex}

\begin{document}

\title{Ultrahigh energy particle collisions near many-dimensional black
holes: general approach }
\author{O. B. Zaslavskii}
\affiliation{Department of Physics and Technology, Kharkov V.N. Karazin National
University, 4 Svoboda Square, Kharkov 61022, Ukraine and\\
Institute of Mathematics and Mechanics, Kazan Federal University, 18
Kremlyovskaya St., Kazan 420008, Russia}
\email{zaslav@ukr.net }

\begin{abstract}
If two particles moving towards a black hole collide near the horizon, their
energy in the centre of mass frame can grow unbounded. This is the so-called
Banados - Silk - West (BSW) effect. Earlier, it was shown that in the 3+1
space-time this effect has a universal nature. We show that for a wide class
of many-dimensional black holes (including, say, the Myers-Perry black hole)
this is also true. The suggested analysis is general and does not require
special properties of the metric like separability of variables for
geodesics, etc.
\end{abstract}

\maketitle

\section{Introduction}

In 2009, Ba\~{n}ados, Silk and West made an interesting observation \cite%
{ban}. It turned out that if two particles collide near the extremal Kerr
black hole, the energy in the centre of mass frame can grow unbounded (the
BSW effect, called after the names of its authors). Later on, the BSW effect
was extended to generic extremal and nonextremal stationary
axially-symmetric black holes, charged static black holes, black holes in
the magnetic field, etc. One of separate issues provoking special interest
is the possibility of generalization to the many-dimensional case. It was
demonstrated in \cite{5d} for the five-dimensional Kerr metric and in \cite%
{d} for many-dimensional Myers-Perry black hole \cite{mp1} that the BSW
effect does exist. It required rather developed studies, especially in \cite%
{d}, where very subtle properties of motion in the Myers-Perry black hole
background were exploited, including the separability of variables \cite{mp1}
- \cite{mp3}.

One may ask: whether the BSW effect for a Myers-Perry black hole arises due
to some special properties of this very space-time? If so, one would be led
to incremental inspection of different models step by step. However, it is
worth reminding that it was shown earlier that for 3+1 black holes the BSW
effect is of universal nature \cite{prd}, \cite{jh} and is due to the
existence of the horizon as such, independently of a particular model.
Therefore, a natural supposition arises that the situation is similar in a
many-dimensional case. In the 3+1 space-time, the BSW effect requires the
existence of so-called critical trajectories with fine-tuned parameters.
Then, it is produced by collision between such a fine-tuned ("critical")
particle and generic ("usual") one. However, both the overall approach and
the formulation of critical condition are so different in \cite{5d}, \cite{d}
and in \ \cite{prd}, \cite{jh}, that the issue of universality is not
obvious in advance.

The aim of the present paper is to show that there exists deep continuity
between 3+1 and higher-dimensional cases. And, the approach to the BSW
effect can be formulated in such a way that, without relying on special
properties of space-time like separability of variables, etc., one can
obtain the critical condition and the effect under discussion as direct
generalization of the 3+1 case.

Throughout the paper we use units in which fundamental constants are $G=c=1$.

\section{Basic equations}

We consider the metric%
\begin{equation}
ds^{2}=-dt^{2}+\frac{dr^{2}}{A(r,\theta )}+\rho ^{2}d\theta
^{2}+\dsum\limits_{i=1}^{n}g_{i}(\{\alpha _{k}\},\theta ,r)d\phi
_{i}^{2}+B(r,\theta )[dt-\dsum\limits_{i}^{n}b_{i}(\{\alpha _{k}\},)d\phi
^{i}]^{2}+\dsum\limits_{i=1}^{n}A_{i}(\{\alpha _{k}\},\theta ,r)d\alpha
_{i}^{2}\text{.}  \label{m1}
\end{equation}

Here, the variable $\theta $ is singled by analogy with the 3+1 case and to
make comparison to the many-dimensional Myers-Perry black hole \cite{d} more
convenient. We assume that all the metric coefficients do not depend on $t$
and $\phi _{i}$, $i=1,...n$. Although the metric (\ref{m1}), with these
restrictions, is not the most general one, it includes physically important
space-times, in particular, the aformentioned Myers-Perry black hole. On the
horizon, $N^{2}=0$.

The metric can be rewritten in the form which is more familar, especially
from the 3+1 context:%
\begin{equation}
ds^{2}=-N^{2}dt^{2}+\dsum\limits_{i=1}^{n}h_{ij}(d\phi ^{i}-\omega
^{i}dt)(d\phi ^{j}-\omega ^{j}dt)+\frac{dr^{2}}{A}+g_{\theta }d\theta
^{2}+\dsum\limits_{i=1}^{n}A_{i}d\alpha _{i}^{2}\text{.}  \label{m2}
\end{equation}%
The relaitonship between two forms of the metric are given by obvious
formulas%
\begin{equation}
-N^{2}+\dsum\limits_{k}h_{ij}\omega ^{i}\omega ^{j}=B-1
\end{equation}%
\begin{equation}
Bb_{i}=h_{ij}\omega ^{j}\text{, }Bb_{i}b_{j}=h_{ij}.\text{ }
\end{equation}%
The inverse metric is equal to

\begin{equation}
g^{00}=-\frac{1}{N^{2}}\text{, }g^{rr}=A\text{, }g^{\theta \theta }=\frac{1}{%
g_{\theta }}\text{, }g^{0i}=-\frac{\omega _{i}}{N^{2}}\text{, }g^{\alpha
_{k}\alpha _{k}}=\frac{1}{A_{k}}\text{,}
\end{equation}%
\begin{equation}
g^{ik}=h^{ik}-\frac{\omega ^{i}\omega ^{k}}{N^{2}}\text{,}
\end{equation}
where $h^{ij}h_{jk}=\delta _{k}^{i}$.

As the metric does not depend on time, the energy $E=-mu_{0}$ is conserved ($%
m$ is a mass), the velocity $u^{\alpha }=\frac{dx^{\alpha }}{d\tau }$, where 
$\tau $ is the proper time. In a similar way, there are conserved angular
momenta corresponding to variables $\phi _{i}$, $mu_{i}=L_{i}$. Then, the
equaitons of motion for geodesics read (dot denotes differentiation with
respect $\tau $):%
\begin{equation}
m\dot{t}=\frac{X}{N^{2}},  \label{t}
\end{equation}%
\begin{equation}
X=E-\dsum\limits_{k}\omega ^{k}L_{k}\text{,}  \label{x}
\end{equation}%
\begin{equation}
m\dot{\phi}^{i}=h^{ij}L_{j}+\frac{\omega ^{i}X}{N^{2}}.  \label{phi}
\end{equation}%
Using the normalization condition $u_{\alpha }u^{\alpha }=-1$, we obtain the
equaiton for the radial component$.$Then, we have%
\begin{equation}
m\dot{r}^{2}=\frac{A}{N^{2}}Z^{2}\text{,}  \label{rz}
\end{equation}%
\begin{equation}
Z^{2}=X^{2}-N^{2}(\dsum\limits_{k}h^{ij}L_{i}L_{j}+m^{2}+\dsum%
\limits_{k}A_{k}\dot{\alpha}_{k}^{2})\text{.}  \label{z}
\end{equation}

\section{Collision of two particles}

If two particles 1 and 2\ collide, one can define their energy in the centre
of mass frame in the point of collision according to%
\begin{equation}
E_{c.m.}^{2}=-(m_{1}u_{1}^{\mu }+m_{2}u_{2}^{\mu })(m_{1}u_{1\mu
}+m_{2}u_{2\mu })=m_{1}^{2}+m_{2}^{2}+2m_{1}m_{2}\gamma \text{,}
\end{equation}%
where the Lorentz factor of relative motion equals%
\begin{equation}
\gamma =-u_{1\mu }u_{2}^{\mu }\text{.}
\end{equation}%
Then, it follows from the equaitons of motion (\ref{t}) - (\ref{z}) that%
\begin{equation}
\gamma =\frac{X_{1}X_{2}-Z_{1}Z_{2}}{m_{1}m_{2}N^{2}}-\dsum%
\limits_{k}h^{ij}L_{1i}L_{2j}-\dsum\limits_{k}A_{k}\dot{\alpha}_{1k}\dot{%
\alpha}_{2k}\text{.}  \label{ga}
\end{equation}

We do not specify the character of motion and only require that $\dot{\alpha}%
_{1k}$ and $\dot{\alpha}_{2k}$ be finite. In particular, there exist
simplest geodesics with all $\alpha _{k}=const$.

Let the point of collision be close to the horizon. In general, this does
not lead to the growth of $\gamma $ since the numerator in $\gamma $
vanishes with the same rate as the denominator. However, there is an
exception. Let us call a particle usual if $X_{H}\neq 0$ and critical if $%
X_{H}=0$ (subscript "H" means that the quantity is calculated on the
horizon). According to (\ref{x}), the condition of criticality reads%
\begin{equation}
E=\dsum\limits_{k}(\omega ^{k})_{H}L_{k}\text{.}  \label{cr}
\end{equation}

Let the critical particle 1 and a usual particle 2 collide near the extremal
horizon. Now, we need the behavior of $X$ near the horizon. We restrict
ourselves by the extremal case, simiularily to what was done in \cite{d} for
Myers-Perry black holes. Then, we can exploit the Taylor expansion near the
horizon: 
\begin{equation}
\omega ^{k}=\omega _{H}^{k}-\omega _{1}^{k}N+O(N^{2})\text{,}  \label{om}
\end{equation}%
where $\omega _{1}$ does not depend on $r$. (For more general discussion of
expansions of this kind for nonextremal and extremal horizons 3+1
dimensional  see Ref. \cite{dirty}. Extension of its results to the
many-dimensional case is straightforward). By substitution into (\ref{x})
and taking eq. (\ref{om}) into account, we obtain that near the horizon%
\begin{equation}
X=CN+O(N^{2})\text{,}
\end{equation}%
where $C=\dsum\limits_{k}\omega _{1}^{k}L_{k}$.

Then, it follows from (\ref{ga}) that near the horizon%
\begin{equation}
\gamma \approx \frac{X_{2}D}{Nm_{1}m_{2}}\text{, }D=C-\sqrt{%
C^{2}-(\dsum\limits_{k}h^{ij}L_{i}L_{j}+m^{2}+\dsum\limits_{k}A_{k}\dot{%
\alpha}_{k}^{2})_{H}}\text{.}
\end{equation}%
Thus for $N\rightarrow 0$ the factor $\gamma $ grows unbounded, and we have
the BSW effect.

\section{Conclusion}

Thus we have shown in a very simple and direct manner that the general
approach to the BSW effect developed earlier for 3+1 black holes \cite{prd}, 
\cite{jh}, works, with some modifications, also in a many-dimensional case.
The key ingredients are preserved: (i) the presence of the horizon, (ii)
classification of particles according to which one of them should be
critical, another one should be usual. In the particular case of the
Myers-Perry black hole, the condition of criticality (\ref{cr}) corresponds
to eq. (2.30) of \cite{d}. As long as we are interested in the BSW effect as
such, there is no need in examining new models one after another and
elucidating complicated and subtle details of particles' motion (which, of
course, can be of interest by themselves). If the metric belongs to the
class (\ref{m1}), (\ref{m2}), the existence of the BSW effect follows from
general grounds.

The general character of our treatment enables us also to use (directly or
with minimum modifications) some other results concerning properties of the
BSW effect. For instance, classification of the possible types of the BSW
scenarios suggested in Sec. 6 of Ref. \cite{jh} now applies and is
insensitive to the number of dimensions. It includes as particular cases the
3+1 Kerr metric (see Sec. IV A of Ref. \cite{kd}) and the Myers-Perry black
hole \cite{d}.

Actually, we showed that the persistence of the BSW effect can be considered
as one more manifestation of universality typical of black holes physics.
This can have interesting physical consequences like, for example,
instability of extremal horizons \cite{d} but this issue requires separate
treatment.

\end{document}